\documentclass[aps,prb,twocolumn,groupedaddress,showpacs,amsmath,amssymb,amsfonts,floats,floatfix,raggedbottom,balancelastpage,dvips]{revtex4}

\usepackage{epsfig} 
\usepackage{psfrag}
\usepackage{braket}
\usepackage{color}
\usepackage{graphicx}
\usepackage{hyperref}
\hypersetup{dvips,
  pdfauthor={Christopher N. Varney},
  pdftitle={Quantum Monte Carlo study of the two-dimensional fermion
    Hubbard model},
  letterpaper,
  colorlinks=true,
  linkcolor=blue,
  citecolor=blue,
  pdfpagemode=UseNone
}
\usepackage{breakurl}

\begin{document}

\title{Quantum Monte Carlo study of the two-dimensional fermion Hubbard Model}
\author{C. N. Varney$^1$, C.-R.~Lee$^2$, 
Z. J. Bai$^3$, S. Chiesa$^1$, M. Jarrell$^4$,
and R. T. Scalettar$^1$} 
\affiliation{$^1$Department of Physics, University of California, Davis,
  California 95616, USA}
\affiliation{$^2$Department of Computer Science, 
National Tsing Hua University, Hsinchu, Taiwan 30013, R.O.C.}
\affiliation{$^3$Department of Computer Science,
University of California, Davis, California 95616, USA}
\affiliation{$^4$Department of Physics and Astronomy,
Louisiana State University, Baton Rouge, LA 70803}

\begin{abstract}
  We report large scale determinant Quantum Monte Carlo calculations
  of the effective bandwidth, momentum distribution, and magnetic
  correlations of the square lattice fermion Hubbard Hamiltonian at
  half-filling.  The sharp Fermi surface of the non-interacting limit
  is significantly broadened by the electronic correlations, but
  retains signatures of the approach to the edges of the first
  Brillouin zone as the density increases.  Finite-size scaling of
  simulations on large lattices allows us to extract the interaction
  dependence of the antiferromagnetic order parameter, exhibiting its
  evolution from weak-coupling to the strong-coupling Heisenberg
  limit.  Our lattices provide improved resolution of the Green's
  function in momentum space, allowing a more quantitative comparison
  with time-of-flight optical lattice experiments.
\end{abstract}


\maketitle

\section{Introduction}
\label{sec:intro}

Originally introduced to explain magnetism and metal-insulator
transitions in solids with strong electronic correlations and narrow
energy bands,\cite{anderson52, hubbard63, hubbard64a, hubbard64b,
  hubbard65} the underlying physics of the fermion Hubbard Hamiltonian
\cite{rasetti91, montorsi92, gebhard97, fazekas99} remains a topic of
considerable discussion.  In two dimensions, when the lattice is doped
away from half-filling, do the fermions condense into a
superconducting state?  If so, what is the symmetry of the pairing
order parameter?\cite{white89, scalapino94,scalapino94b, maier05,
  scalapino07} Do charge inhomogeneities (stripes and checkerboards)
emerge, and what is their interplay with magnetic and superconducting
orders?\cite{millis98, white00, moreo05, chang08}

In contrast to this uncertainty concerning the properties of the doped
lattice, the qualitative behavior at half-filling (one fermion per
site) is much more well understood.  The interaction strength $U$
causes both the development of long range antiferromagnetic order
(LRAFO) and insulating behavior.  Even so, there are still some
remaining open fundamental questions, for example, in the precise way
in which the model evolves from the weak-coupling to strong-coupling
limits, especially in two dimensions.

At weak-coupling, one pictures the insulating behavior to arise from a
Fermi-surface instability which drives LRAFO and a gap in the
quasiparticle density of states.  On the other hand, at strong
coupling the insulating behavior is caused by Mott physics and the
suppression of electron mobility to avoid double occupancy.  These
points of view are clearly linked, however, since for large $U/t$ the
Hubbard Hamiltonian has well defined local moments and maps onto the
antiferromagnetic Heisenberg model with exchange constant $J =
4t^2/U$.\cite{reger88}

Developing an analytic theory which bridges these viewpoints
quantitatively is problematic.  Hartree Fock (HF) theory provides one
simple point of view, but predicts LRAFO at finite temperatures in two
dimensions, in violation of the Mermin-Wagner theorem.  In fact, even
in higher dimension when the N\'eel tempertaure $T_{\rm N}$ can be
nonzero, HF theory predicts $T_{\rm N} \propto U$ instead of the
correct $T_{\rm N} \propto J = 4t^2/U$.  Sophisticated approaches such
as the self-consistent renormalized theory,\cite{moriya85,moriya00}
the fluctuation-exchange approximation,\cite{bickers89} and
two-particle self-consistent theory\cite{vilk97} obey the
Mermin-Wagner theorem and provide a good description of the Hubbard
Hamiltonian at weak-coupling, but fail for large $U/t$.  A recent
approach \cite{borejsza04} based on the mapping to the nonlinear sigma
model \cite{schultz95} has made some progress in connecting the two
regimes.

The need to pin down the behavior of the two-dimensional (2D) half
filled Hubbard model more quantititively, in a way which links the
weak-coupling and strong-coupling limits, is particularly germane at
present with the achievement of cooling and quantum degeneracy in
ultracold gases of fermionic atoms.\cite{demarco99, truscott01,
  ohara02, kohl05, kohl06, jordens08} Such systems offer the prospect
of acting as an ``optical lattice emulator'' (OLE) of the fermion
Hubbard model, allowing a precise comparison of experimental and
theoretical phase diagrams which is difficult in the solid state,
where the (single band) Hubbard Hamiltonian provides only a rather
approximate depiction of the full complexity of the atomic orbitals.
Obviously, the achievement of this goal is one which requires accurate
computations.  A particular issue in the field of OLE concerns whether
the temperature dependence of the double occupancy rate changes sign
during the course of the evolution from weak to strong
coupling.\cite{werner05}

It is the intent of this paper to present considerably improved
results for the effective bandwidth, momentum distribution, and
magnetic correlations of the square lattice fermion Hubbard
Hamiltonian.  We will employ the determinant quantum Monte Carlo
(DQMC) method, which provides an approximation-free solution of the
model, on lattices large enough to use finite-size scaling to, for
example, reliably extract the antiferromagnetic order parameter as a
function of interaction strength. There is a considerable existing
body of QMC studies of the two-dimensional half-filled Hubbard model,
both on finite lattices and in infinite dimension.  A partial list
includes Refs.~\onlinecite{maier05},~\onlinecite{chang08}, and
\onlinecite{vollhardt93, bulut94, haas95, georges96, preuss97,
  grober00, maier05rmp, dare07}.

\section{Model and Computational Methods}
\label{sec:model_comp_methods}

\begin{figure}[tb]
  \includegraphics*[height=\columnwidth,angle=-90,viewport=0 0 612 790]{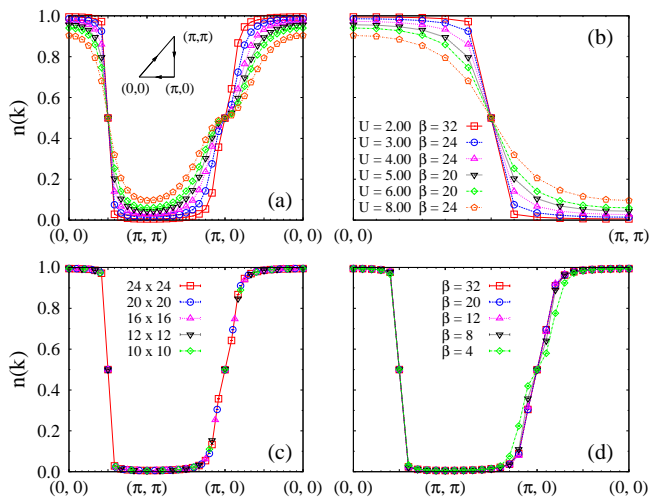}
  \caption{(Color online)  
    [(a) and (b)] The momentum distribution, Eq.~(\ref{eq:momdistr}), is
    shown for interaction strengths $U$ ranging from $U=2t$ (one
    quarter the bandwidth) to $U=W=8t$.  A sharp Fermi surface is seen
    at weak-coupling as the momentum cuts across the Fermi surface at
    ${\bf k}=(\pi/2,\pi/2)$.  Larger $U$ broadens $n({\bf k})$
    considerably.  The occupation becomes substantial outside the
    nominal Fermi surface.  Panel (a) shows the full BZ, while panel
    (b) provides higher resolution for the portion of the cut
    perpendicular to the Fermi surface at $(\pi/2,\pi/2)$.  (c) At $U
    = 2t$ and $\beta t = 32$, $n({\bf k})$ has only a weak lattice size
    dependence, apart from the better resolution as $L$ increases.
    (d) For $U = 2t$ on a $20 \times 20$ lattice, $n({\bf k})$ is
    converged to its low temperature value once $T < t/8$.  (By
    contrast, the spin correlations reach their ground state values
    only at considerably lower $T$.)   
    \label{fig:GqdiffU}
  }
\end{figure}

The fermion Hubbard Hamiltonian,
\begin{align}
  \begin{aligned}
    H = &-t \sum_{\braket{{\bf i \, j}} \sigma} \left( c^\dag_{{\bf j}
        \sigma} c^{\phantom \dag}_{{\bf i} \sigma} + c^\dag_{{\bf i}
        \sigma} c^{\phantom \dag}_{{\bf j} \sigma} \right)\\
    &+ U \sum_{\bf i} \left( n_{{\bf i} \uparrow}^{\phantom \dag} -
      \frac{1}{2} \right) \left( n_{{\bf i} \downarrow}^{\phantom
        \dag} - \frac{1}{2} \right) - \mu \sum_{{\bf i} \sigma}
    n_{{\bf i} \sigma}^{\phantom \dag},
  \end{aligned}
  \label{eqn:hubham}
\end{align}
describes a set of itinerant electrons, represented by $c^{\phantom
  \dag}_{{\bf j}\sigma} (c^\dag_{{\bf j}\sigma})$, the annihilation
(creation) operators at lattice site ${\bf j}$ and spin $\sigma$. The
corresponding number operator $n_{{\bf j} \sigma} = c^\dag_{{\bf
    j}\sigma}c^{\phantom \dag}_{{\bf j}\sigma}$.  The first term
represents the hopping (kinetic energy) of the electrons.  We will
choose the parameter $t = 1$ to set our unit of energy.  The
non-interacting bandwidth $W=8t$.  $U$ is the on-site repulsion of
spin-up and spin-down electrons occupying the same lattice site, and
$\mu$ is the chemical potential which controls the particle density.
We will mostly be interested in the properties of the model on $N = L
\times L$ square lattices at half-filling (the number of particles is
equal to the number of lattice sites) which occurs at $\mu = 0$ with
our particle-hole symmetric choice of the representation of the
interaction term.

We will also focus exclusively on the case of the square lattice.
This particular geometry has several interesting features.  The
half-filled square lattice Fermi surface exhibits perfect nesting, and
the density of states is (logarithmically) divergent.  As a
consequence, the antiferromagnetic and insulating transitions occur
immediately for any nonzero value of the interaction strength $U$,
instead of requiring a finite degree of correlation, as is more
generically the case.

Our DQMC algorithm is based on Ref.~\onlinecite{blankenbecler81} and
has been refined by including ``global moves'' to improve ergodicity
\cite{scalettar91} and ``delayed updating'' of the fermion Green's
function,\cite{jarrell09} which increases the efficiency of the linear
algebra.  Details concerning this new code are available at
Ref.~\onlinecite{quest09}.  Some other approaches to fermion Hubbard
model simulations are contained in Refs.~\onlinecite{maier05rmp} and
\onlinecite{hubsims2, vanhoucke08, prokofev08}.

\section{Single-Particle Properties}
\label{sec:hoppingandgreensfunction}

We begin by showing single-particle properties.  The momentum
distribution $n({\bf k}) = \frac{1}{2}\sum_\sigma \braket{
  c^{\dagger}_{{\bf k}\sigma} c^{\phantom \dagger}_{{\bf k}\sigma}}$
is obtained directly in DQMC via Fourier transform of the equal-time
Green's function $G_{\bf j i} = \braket{c_{{\bf
      j}\sigma}^{\phantom\dag} c_{{\bf i}\sigma}^\dag}$
\begin{align}
  n({\bf k})=1-\frac{1}{2N}\sum_{{\bf i,j},\sigma}e^{i{\bf k \cdot (j-i)}} 
  \braket{ c^{\phantom \dagger}_{\bf j\sigma}c^\dagger_{\bf
    i\sigma} }.
  \label{eq:momdistr}
\end{align}
At $U=0$ and at half-filling, $n({\bf k})=1(0)$ inside (outside) a
square with vertices $(\pi,0)$, $(0,\pi)$, $(-\pi,0)$, and $(0,-\pi)$
within the Brillouin zone (BZ).  In Fig.~\ref{fig:GqdiffU}(a), we show
$n({\bf k})$ around the complete BZ, while Fig.~\ref{fig:GqdiffU}(b)
focuses on the region near the Fermi-surface point $(\pi,2,\pi/2)$.
Interactions broaden the $U=0$ Fermi surface considerably.
Figure~\ref{fig:GqdiffU}(c) shows that data for different lattice
sizes fall on the same curve.  Smearing due to finite temperature
effects is seen in Fig.~\ref{fig:GqdiffU}(d) to be small below $T = t
/ $ ($\beta t = 8$).

\begin{figure*}[tb]
  \includegraphics[width=0.7\textwidth,viewport=0 150 612 660]{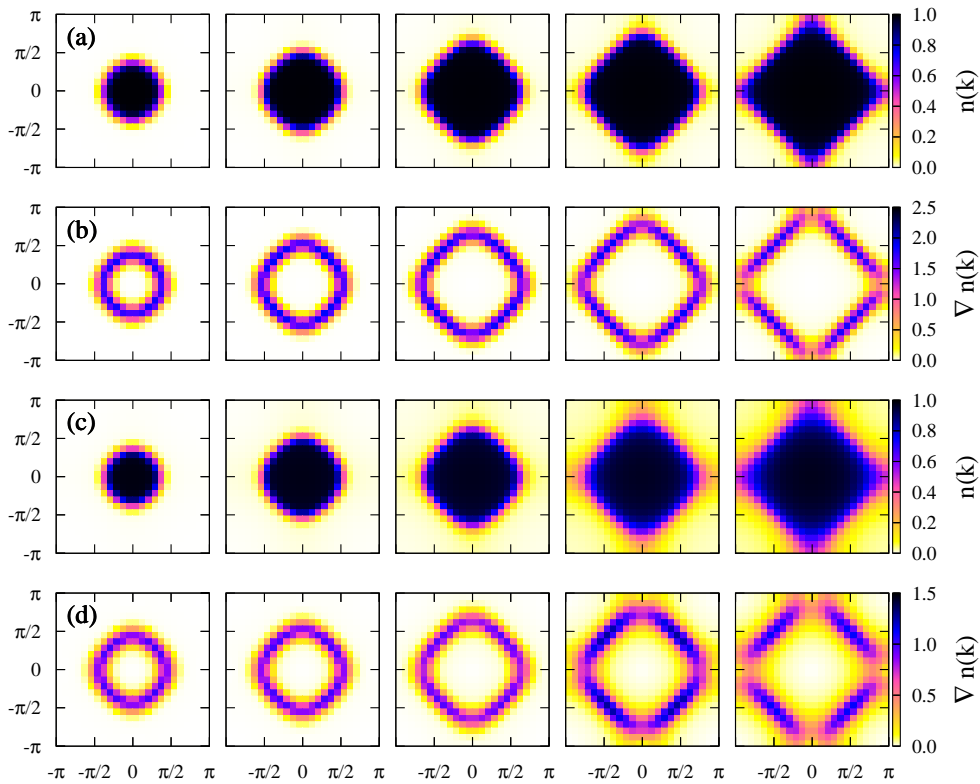}
  \caption{(Color online)
    Color contour plot depiction of the momentum distribution $n({\bf k})$
    and its gradient $\nabla n({\bf k})$.  (a) Left to right,
    $n({\bf k})$ at weak-coupling $U=2t$ and fillings $\rho=0.23$,
    $0.41$, $0.61$, $0.79$, and $1.0$.  (b) $\nabla n({\bf k})$ for
    the same parameters.  [(c) and (d)] Intermediate coupling $U=4t$
    and fillings $\rho = 0.21$, $0.41$, $0.59$, $0.79$, and $1.0$.
    The increased breadth of the Fermi surface with interaction
    strength is evident. The lattice size = $24\times24$ and inverse
    temperature $\beta t = 8$ except at $U = 4t$ and fillings $\rho =
    0.59$ and $0.79$, where the sign problem restricts the simulation
    to inverse temperatures $\beta t = 6$ and $4$, respectively.
    \label{fig:greenexpt}
  }
\end{figure*}

Recent optical lattice experiments \cite{kohl05} have imaged this
Fermi surface for a three-dimensional cloud of fermionic $^{40}$K
atoms prepared in a balanced mixture of two hyperfine states which act
as the Hubbard Hamiltonian spin degree of freedom.  In
Fig.~\ref{fig:greenexpt} we show a sequence of color contour plots for
different densities at weak and intermediate couplings, $U/t=2,4$.  As
in the experiments, and in agreement with Fig.~\ref{fig:GqdiffU}, the
Fermi surface may still be clearly discerned, and evolves from a
circular topology at low densities into the rotated square as the BZ
boundaries are approached.  Because of the sign
problem\cite{hirsch85,loh90} which occurs in the doped system, the
temperatures shown in the figure are rather higher than those used in
Fig.~\ref{fig:GqdiffU} at half-filling.
\begin{figure}[hptb]
  \includegraphics*[height=0.9\columnwidth,angle=-90,viewport=0 0 612 790]{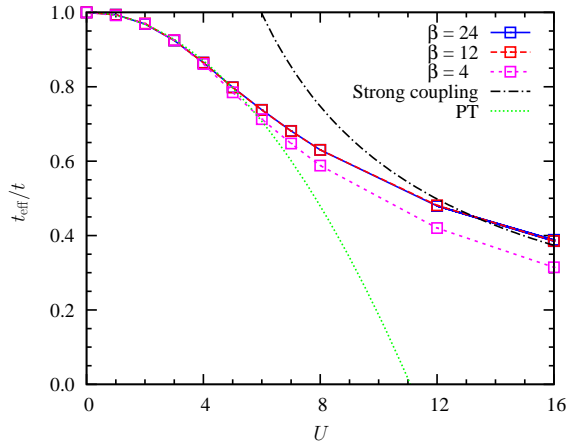}
  \caption{(Color online)
    As the interaction energy $U$ increases, the kinetic energy
    declines.  Here we show $t_{\rm eff}/t$, the ratio between the
    expectation value of $\braket{ c^{\dagger}_{{\bf j}+\hat x \,
        \sigma} c^{\protect \phantom \dagger}_{{\bf j} \, \sigma} }$
    at $U$ with its value at $U=0$, for a $10\times10$ lattice.
    Strong-coupling and perturbative graphs are also shown for $\beta
    t = 12$. 
    \label{fig:teff}
  }
\end{figure}
Another single-particle quantity of interest is the effective hopping,
\begin{align}
  \frac{t_{\rm eff}}{t} &=
  \frac{
    \braket{
      c^{\dagger}_{{\bf j}+\hat x \, \sigma} 
      c^{\phantom \dagger}_{{\bf j} \, \sigma} 
      + c^{\dagger}_{{\bf j} \, \sigma} 
      c^{\phantom \dagger}_{{\bf j}+\hat x \, \sigma} 
    }_{U}
  }
  {
    \braket{
      c^{\dagger}_{{\bf j}+\hat x \, \sigma} 
      c^{\phantom \dagger}_{{\bf j} \, \sigma} 
      + c^{\dagger}_{{\bf j} \, \sigma} 
      c^{\phantom \dagger}_{{\bf j} +\hat x\, \sigma} 
    }_{U=0}
  },
\label{eqn:teff}
\end{align}
which measures the ratio of the kinetic energy at finite $U$ to its
non-interacting value.  As the electron correlations grow larger,
hopping is increasingly inhibited, and $t_{\rm eff}$ is diminished.
In Fig.~\ref{fig:teff}, we show a plot of this ratio as a function of
$U$ for a $10 \times 10$ lattice.  Note that despite the insulating
nature of the system, the effective hopping is nonzero and does not
serve as an order parameter for the metal-insulator transition.
Indeed, $t_{\rm eff}$ is responsible for the superexchange interaction
which drives antiferromagnetic order.  The effective hopping can be
evaluated analytically at small and large $U$ \cite{white89}.  The
DQMC data interpolates between these two limits.

\section{Magnetic Correlations}
\label{sec:magnetism}

We turn now to two-particle properties, focusing on the magnetic
behavior.  The real-space spin-spin correlation function is defined
as
\begin{align}
  C({\bf l}) &= \braket{(n_{{\bf j} + {\bf l} \uparrow} - n_{{\bf j} +
      {\bf l} \downarrow})(n_{{\bf j} \uparrow} - n_{{\bf j}
      \downarrow})}
  \label{eq:spinspin}
\end{align}
and measures the extent to which the $z$ component of spin on site ${\bf
j}$ aligns with that on a site a distance ${\bf l}$ away.  Although
defined in Eq.~(\ref{eq:spinspin}) using the $z$ direction, $C({\bf l})$
is rotationally invariant and in fact, we measure all three components
to monitor ergodicity in our simulations and average over all
directions to provide an improved estimator for the magnetic properties.

\begin{figure}[t]
  \includegraphics*[height=0.9\columnwidth,angle=-90,viewport=0 0 612 792]{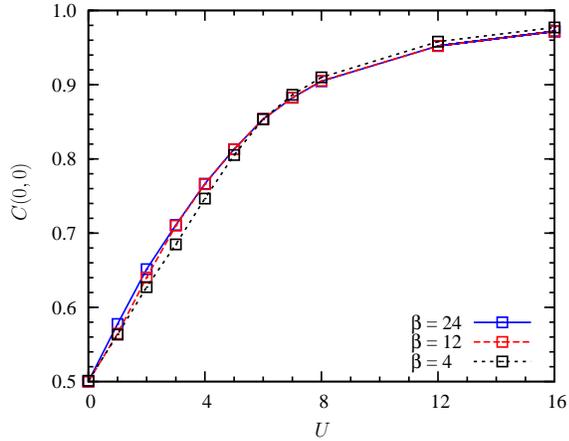}
  \caption{(Color online)
    The local moment $\langle m^2 \rangle$ is the zero spatial
    separation value of the spin-spin correlation function $C(
      0,0)$. In the non-interacting limit $\langle m^2 \rangle = \frac12$.
    As the interaction energy $U$ increases, $\langle m^2 \rangle$
    approaches 1, indicating the complete absence of double occupancy
    and a well-formed moment on each site.  }
    \label{fig:localmoment}
\end{figure}
\begin{figure}[b]
  \includegraphics*[height=0.9\columnwidth,angle=-90,viewport=0 0 612 792]{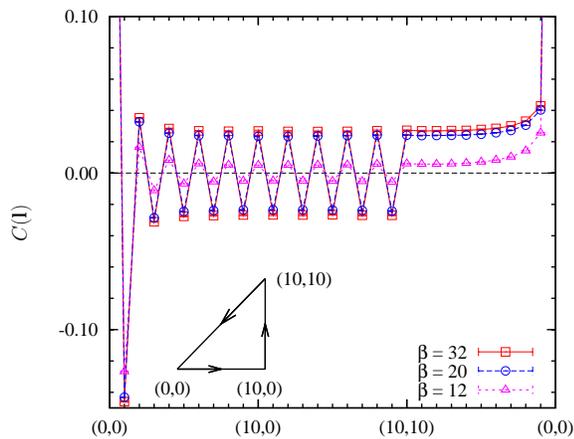}
  \caption{(Color online)
    Comparison of the equal-time spin-spin correlation function
    $C({\bf l})$ on a $20\times20$ lattice with $\braket{n} = 1$ and
    $U = 2t$ for inverse temperatures $\beta t = 12, 20,$ and
    $32$. The horizontal axis follows the triangular path on the 
    lattice shown in the inset. Anti-ferromagnetic correlations are
    present for all temperatures, and saturation is visible at $\beta
    t = 32$. 
    \label{fig:corr3}
  }
\end{figure}
The local moment $\braket{ m^2 } = C(0,0) = \braket{ (n_{{\bf j}
    \uparrow} - n_{{\bf j} \downarrow})^2 }$ is the zero separation
value of the spin-spin correlation function.  The singly occupied
states $\ket{\,\uparrow\,}$ and $\ket{ \, \downarrow \, }$ have $
\braket{m^2} = 1$ while the empty and doubly occupied ones $\ket{ \, 0
  \, }$ and $\ket{ \, \uparrow \downarrow \, }$, have $ \braket{m^2} =
0$.  In the non-interacting limit, at half-filling, each of the four
possible site configurations is equally likely.  Hence the average
moment $\braket{m^2} = \frac{1}{2}$.
\begin{figure}[t]
  \includegraphics*[height=0.9\columnwidth,angle=-90,viewport=0 0 612 792]{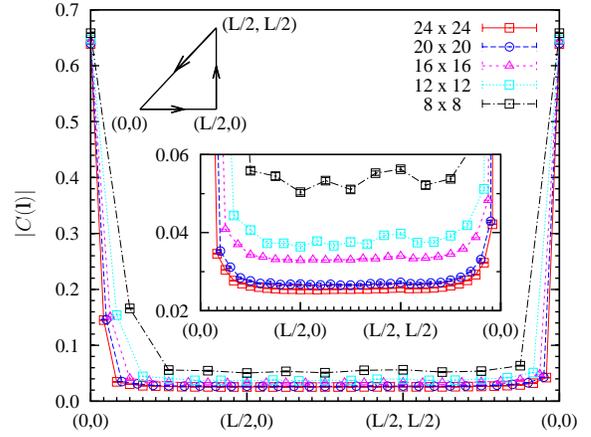}
  \caption{(Color online)
    Comparison of the absolute value of the equal-time spin-spin
    correlation function $C({\bf l})$ at $U = 2t$ and $\beta t = 32$
    for $L \times L$ lattices with $L = 8, 12, 16, 20$, and $24$. The
    inset is a close up view of the long-range correlations. 
    \label{fig:corr2}
  }
\end{figure}
\begin{figure}[b]
  \includegraphics*[height=0.9\columnwidth,angle=-90,viewport=0 0 612 792]{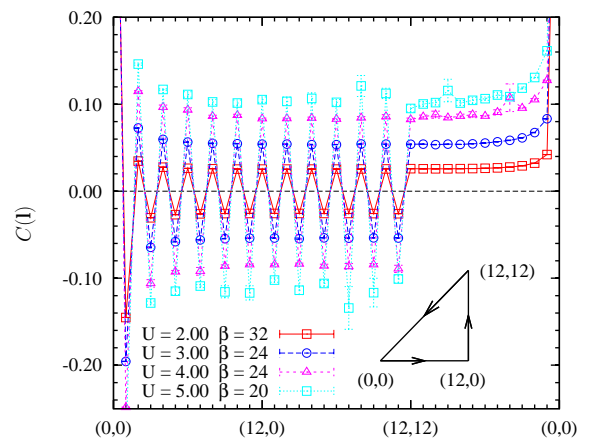}
  \caption{(Color online)
    The equal-time spin-spin correlation function $C({\bf l})$ on a
    $24 \times 24$ lattice with $\braket{n} = 1$. Data are shown for
    various $U$ at low temperatures. Antiferromagnetic correlations
    are enhanced for larger values of $U$. The increase in
    statistical fluctuations with interaction strength in the DQMC
    method is evident. 
    \label{fig:corr}
  }
\end{figure}
The on-site repulsion $U$ suppresses the doubly occupied configuration
and hence also the empty one, if the total occupation is fixed at one
fermion per site.  Ultimately, charge fluctuations are completely
eliminated, $\langle m^2 \rangle \rightarrow 1$ and the Hubbard model
maps onto the spin-$\frac12$ Heisenberg Hamiltonian.  This is
illustrated in Fig.~\ref{fig:localmoment} for a $10 \times 10$
lattice.  By the time $U=W=8t$, the local moment has attained 90\% of
its full value.  Thermal fluctuations also inhibit local moment
formation but the data shown for different temperatures in
Fig.~\ref{fig:localmoment} indicate they are mostly eliminated by the
time $T$ decreases below $t/12=W/96$.

Local moments provide an intuitive picture of the onset of long-range
correlation in the strong-coupling regime. They first form on the
temperature scale $U$, which acts to eliminate double occupancy, and
then, at yet lower $T$, they order via antiferromagnetic exchange
interaction with $J=4t^2/U$. In contrast with this situation,
weak-coupling correlations are better described as arising from the
instability of the Fermi gas against formation of a spin-density wave,
a peculiarity of the square lattice, suggesting an ordering
temperature proportional to $U$.
\begin{figure}[t]
  \includegraphics*[height=\columnwidth,angle=-90,viewport=0 0 612 792]{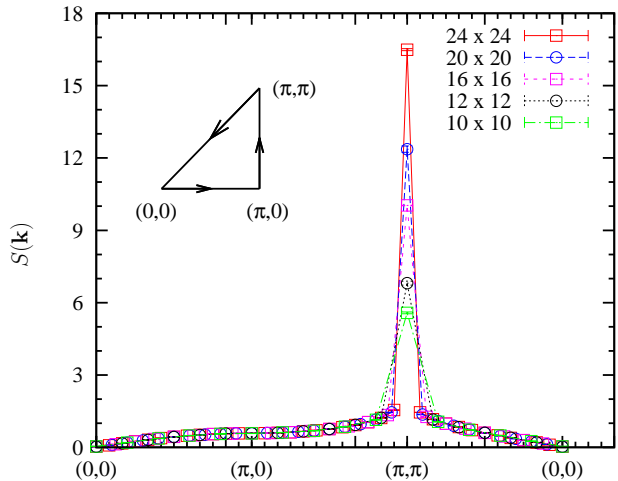}
   \caption{(Color online)
    The magnetic correlation function $S({\bf k})$ for $U = 2t$ and
    $\beta t = 32$. The horizontal axis traces out the triangular
    path shown in the inset. The function is sharply peaked at ${\bf
      k} = {\bf Q} \equiv (\pi,\pi)$. 
    \label{fig:struct}
  }
\end{figure}

Figure~\ref{fig:corr3} shows the spin-spin correlation function in the
latter regime ($U=2$) for a $20\times 20$ lattice at $\beta t=12$,
$20$ and $32$. The correlations extend over the entire lattice even at
$\beta t=12$, i.e., the correlation length has become comparable to
the system size already at this temperature. The values of $C({\bf
  l})$ continue to grow as $T$ is increased further, saturating at
$\beta t\simeq 32$.  This observation disproves the commonly held idea
that on finite clusters, the order parameter stop growing after the
correlation length exceeds the linear size of the system. Such
saturation happens at a much lower temperature, only after thermal
fluctuations have been largely eliminated.

A comparison of $\lvert C({\bf l}) \rvert $ for $U=2$ and different
lattice sizes is given in Fig.~\ref{fig:corr2}, where data for $L = 8$
up to $L = 24$ are plotted and we have taken the absolute value to
make the convergence with $L$ clearer.  We have fixed $\beta t = 32$
so the spin correlations have reached their asymptotic low-temperature
values.  As expected, the smallest lattice sizes ($8\times8$)
overestimate the tendency to order, with $\lvert C({\bf l}) \rvert$
significantly larger than values for larger $L$.  However, by the time
$L = 20$ the finite-size effects are small.

We next compare the spin-spin correlation function for various $U$ at
low temperatures on a $24 \times 24$ lattice in Fig.~\ref{fig:corr}.
Long-range order is present at all interaction strengths.  For each
$U$, we have chosen temperatures such that the ground state has been
reached for this lattice size.  Since statistical fluctuations
increase significantly with $U$ and with $\beta$ in DQMC, it is
advantageous not to simulate unnecessarily cold systems.  As discussed
above, such temperature should increase with $U$ in the weak-coupling
regime and scale proportionally to $1/U$ in the strong-coupling
one. We indeed find the highest saturation temperatures in the
intermediate regime, at $U / t \simeq 4$.

The magnetic structure factor $S({\bf k})$ is the Fourier transform of
the real-space spin-spin correlation function $C({\bf l})$,
\begin{align}
  S({\bf k}) = \sum_{\bf l} e^{i {\bf k} \cdot {\bf l}} C({\bf l}),
\label{eqn:structfac}
\end{align}
were $S({\bf k})$ is plotted in Fig.~\ref{fig:struct} as a function of
${\bf k}$ for several lattice sizes with $U = 2t$ and $\beta t = 32$.
$S({\bf k})$ is small and lattice size independent away from the
ordering vector ${\bf k}={\bf Q}\equiv(\pi,\pi)$.  The sharp peak at
${\bf Q}$ emphasizes the antiferromagnetic nature of the correlations
on a half-filled lattice.

\begin{figure}[t]
  \includegraphics*[height=\columnwidth,angle=-90,viewport=0 0 612 792]{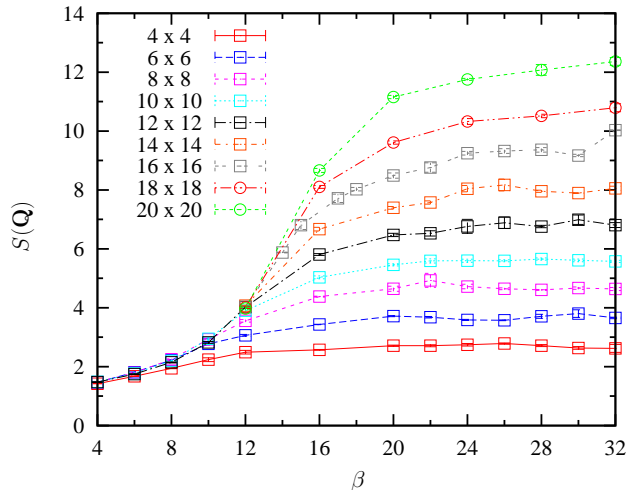}
  \caption{(Color online)
    The antiferromagnetic structure factor $S({\bf Q})$ as a function
    of inverse temperature at $U = 2t$ for $L \times L$ lattices with
    $L = 4, 6, 8, 10, 12, 14, 16, 18$, and $20$.
    \label{fig:struct_v_beta}
  }
\end{figure}

In order to understand the implications of the lattice size dependence
at the ordering vector in Fig.~\ref{fig:struct}, we show in
Fig.~\ref{fig:struct_v_beta} the antiferromagnetic structure factor
for $U = 2t$ as a function of inverse temperature for various $L$.  As
expected, as $L$ increases, a larger value of $\beta$ is required to
eliminate the low-lying spin-wave excitations and to saturate the
structure factor to its ground state value.

\begin{figure}[t]
  \includegraphics*[height=\columnwidth,angle=-90,viewport=0 0 612
  792]{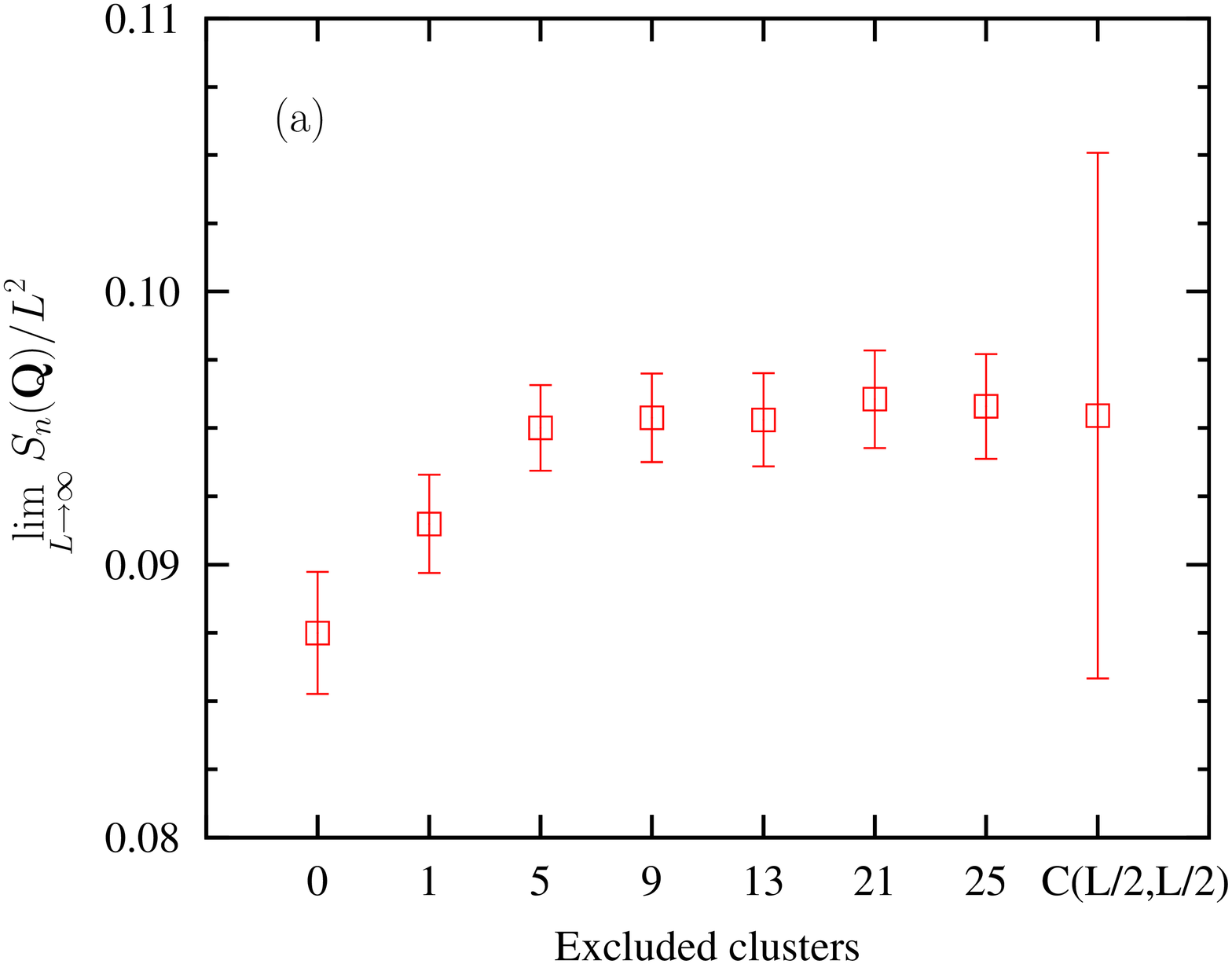}
  \includegraphics*[height=\columnwidth,angle=-90,viewport=0 0 612
  792]{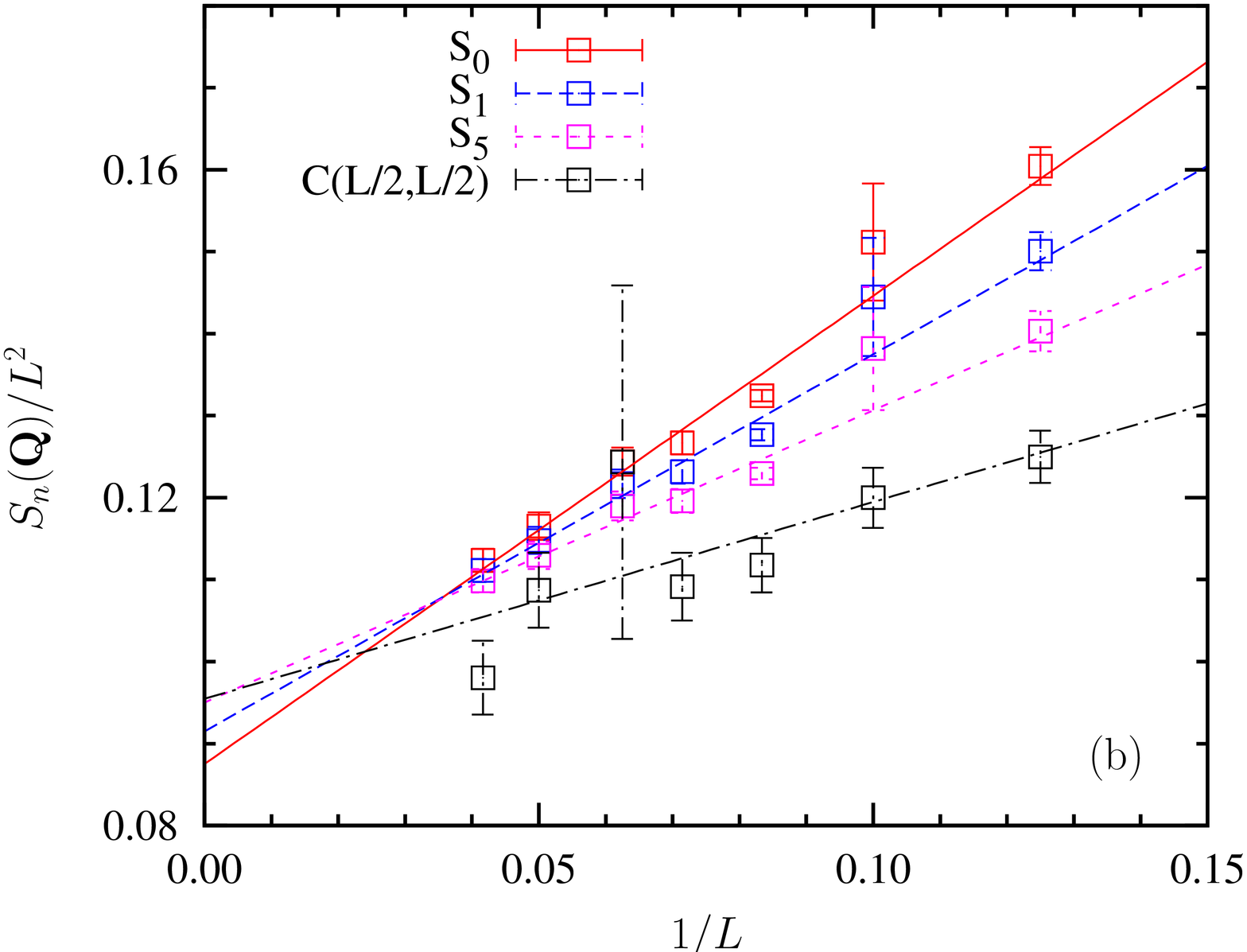}
  \caption{(Color online)
    Scaling results for $U = 5t$ and $\beta t = 20$.
    (a) Convergence of the extrapolated value of $S_n({\bf Q})$ as a
    function of the size of the excluded cluster. See
    Eq.~(\ref{eqn:excvol}) for the definition of $S_n$. A large fraction
    of the $1/L^2$ bias is removed without loss in
    precision. $C(L/2,L/2)$ is plagued by a much larger error bar. (b)
    Scaling of $S_n({\bf Q})$ for $n = 0$, $1$, and $5$ and $C(L/2,L/2)$
    as a function of the inverse linear lattice size. The
    extrapolation was performed via a linear least-squares fit in all
    cases. 
    \label{fig:excl}
  }
\end{figure}

It is seen from Eq.~(\ref{eqn:structfac}) that $S({\bf Q})$ will grow
linearly with the number of sites $N=L^2$ if there is long-range
antiferromagnetic order.  Huse\cite{huse88} has used spin-wave theory
to work out the first correction to this scaling,
\begin{align}
  \frac{S({\bf Q})}{L^2} &= \frac{m_{\rm af}^2}{3} + \frac{a}{L}.
  \label{eqn:husescalingS}
\end{align}
Here $m_{\rm af}$ is the antiferromagnetic order parameter.  $m_{\rm af}$
can also be extracted from the spin-spin correlation function between
the two most distant points on a lattice, $C(L/2,L/2)$, with a similar
spin-wave theory correction,
\begin{align}
  C(L/2, L/2) &= \frac{m_{\rm af}^2}{3} + \frac{b}{L}.
  \label{eqn:husescalingC}
\end{align}
\begin{figure}[t]
  \includegraphics*[height=\columnwidth,angle=-90,viewport=0 0 612 792]{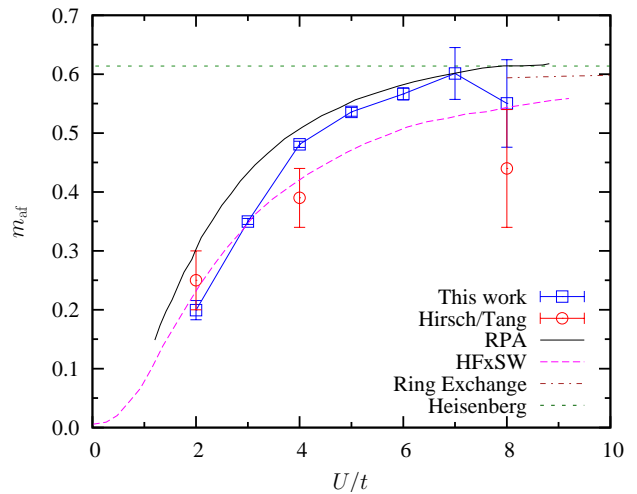}
  \caption{(Color online)
    Order parameter $m_{\rm af}$ as a function of the interaction
    strength $U$. Earlier DQMC values (Ref.~\onlinecite{hirsch89}) are
    circles and Hartree-Fock theory scaled by the Heisenberg result
    from Ref.~\onlinecite{hirsch85} are shown as a line with long
    dashes. The line with short dashes and the dashed and dotted line
    indicate the strong-coupling Heisenberg limits from QMC
    (Ref.~\onlinecite{sandvik97} and spin-only low-energy theory with
    ring exchange (Ref.~\onlinecite{delannoy05}), respectively. Also
    shown is the RPA calculation of Ref.~\onlinecite{schrieffer89}
    (solid line).
    \label{fig:order}
  }
\end{figure}
We expect that the correction $b<a$ since the structure factor
includes spin correlations at short distances which markedly exceed
$m_{\rm af}^2$, in addition to the finite lattice effects at larger
length scales. For similar reasons $S$ is expected to show larger
corrections to the asymptotic $1/L$ scaling behavior than $C$. Part of
the origin of these corrections is trivial and evident from
Fig.~\ref{fig:corr2}: $S$ is the average of quantities as different as
$C(0,0)$ and $C(L/2,L/2)$.  In the large $L$ limit, $S$ and its $1/L$
finite-size error are dominated by the contribution of the large
distance correlations but for small $L$ a bias roughly proportional to
\begin{align}
  \frac{C(0,0)-C(L/2,L/2)}{L^2}
  \label{eq:bias}
\end{align}
is clearly present. This bias is larger for small $U$ since the
numerator in Eq.~(\ref{eq:bias}) gets smaller with increasing $U$ and
saturates to the value of the Heisenberg model at $U/t\simeq 8$.  On
the other hand, the error bars on $S$ are often significantly smaller
than on $C$, a fact that is certainly advantageous in the final
finite-size scaling analysis.

A measure of magnetic order that incorporates the extended linearity
of $C$ and the better statistical property of $S$ is given by
\begin{align}
  S_n({\bf Q}) = \frac{L^2}{L^2-n} \, \sum_{{\bf l},l>l_c} e^{i {\bf Q}
    \cdot {\bf l}} C({\bf l}), 
\label{eqn:excvol}
\end{align}
where $n$ is the number of distances shorter than
$l_c$. Equation~(\ref{eqn:excvol}) is nothing but the interpolation
between $S$, corresponding to $n=0$, and $C$, the case of
$n=L^2-1$. Figure \ref{fig:excl}(a) shows how the $L\rightarrow\infty$
extrapolation evolves by increasing $l_c$. When $l_c$ is small the
linear extrapolation is significantly biased by the small-$L$
results. As $l_c$ increases one reaches statistical convergence
already for $l_c=1$ (corresponding to $n=5$, the cluster formed by the
origin and the nearest neighbors) with minimal loss of statistical
precision. That $l_c=1$ is all is needed to reach statistical
convergence is also manifest in Fig.~\ref{fig:corr2} where $C({\bf
  l})$ drops to almost a constant beyond this value.

In Fig.~\ref{fig:excl}(b) we show $S_n({\bf Q}) / L^2$ versus inverse
linear lattice size $1 / L$ for $U = 5t$ and $n = 0$, $1$, and $5$
since as shown in Fig.~\ref{fig:excl}(a), there is no real gain in
accuracy by excluding larger subclusters.  The inverse temperatures is
$\beta t = 20$, so that $S$ has reached its zero-temperature value
regardless of $L$.  We have repeated this finite-size scaling analysis
for couplings $U/t=2$,$3$,$4$,$6$,$7$, and $8$ and extrapolated to
infinite $L$ using a linear least-square fit in $1/L$. In
Fig.~\ref{fig:order}, we show the resulting antiferromagnetic order
parameter $m_{\rm af}$ as a function of $U/t$ employing the same
normalization convention of the other Hubbard model studies reported
in this figure.  The early DQMC values obtained by Hirsch and
Tang,\cite{hirsch89} which are consistent with the ones obtained here,
are shown.

Figure~\ref{fig:order} also summarizes a number of the available
analytic treatments.  The line with long dashes is the result of
Hartree-Fock theory scaled by the Heisenberg result at strong
coupling.\cite{hirsch85} The solid line is the
random-phase-approximation (RPA) treatment in which the
single-particle propagators in the usual RPA sum are also dressed by
the one-loop paramagnon correction to their
self-energy.\cite{schrieffer89} Also shown (line with dots and dashes)
are the results of a spin-only low energy theory \cite{delannoy05}
which includes not only the usual Heisenberg $J = 4t^2/U$ but also all
higher order (e.g., ring exchange) terms up to $t^4/U^3$.  Finally,
the line with short dashes is the Heisenberg value determined by
Sandvik.\cite{sandvik97}

$\,$\vspace{-20pt}

\section{Summary}
\label{sec:summary}

In this paper we have presented the results of determinant quantum
Monte Carlo calculations for the magnetic properties of the
half-filled square lattice Hubbard Hamiltonian.  DQMC allows us to
bridge the weak-coupling and strong-coupling regimes with a single
methodology and a particular outcome of our work has been the
calculation of the antiferromagnetic order parameter in the ground
state as a function of $U / t$. We expect these values will be useful
in validating OLE experiments on the fermion Hubbard model.

By using an improved DQMC code, we have been also able to provide
results on larger lattices than those originally
explored.\cite{white89} This not only has allowed us to do more
accurate finite-size scaling for the order parameter but we also
obtain considerably better momentum resolution and hence a description
of the Green's function $G({\bf k})$ which also offers the prospect of
improved contact with time-of-flight images from optical lattice
emulators.\cite{ohara02,kohl05,kohl06,jordens08}

This study demonstrates a significantly improved capability to
simulate interacting fermion systems, driven by more powerful hardware
as well as algorithmic advances.  Systems of 500 sites (fermions) can
now be handled on a modest cluster of desktop computers.  Larger
system simulations can easily be contemplated using more powerful
hardware and would scale as the cube of the number of particles, in
the absence of the sign problem.  This remaining sign problem
bottleneck prevents the study of the densities of most interest to
high-temperature superconductivity, i.e., dopings of $5-15$ \% away
from half-filling and motivates the interest in analog computation
for the Hubbard Hamiltonian.\cite{feynman82} It should be noted,
however, that the sign problem can be rather modest for other
densities, e.g.,  quarter filling, where we now have the
capability to undertake large scale studies.

\section*{ACKNOWLEDGEMENTS}
\label{sec:acknowledgements}
Research supported by the DOE SciDAC and SSAAP Programs (Grants
No. DOE-DE-FC0206ER25793 and No. DOE-DE-FG01-06NA26204), and by ARO
under Award No. W911NF0710576 with funds from the DARPA OLE
Program. We thank G. Lightfoot for useful input. RTS thanks the Aspen
Center of Physics for its program on QUantum Simulation/Computation.

\bibliography{hub2d}

\begin{thebibliography}{10}
\providecommand{\selectlanguage}[1]{\relax}

\bibitem{anderson52}
P.~W. Anderson.
\newblock Phys. Rev. \textbf{86}, 694 (1952).

\bibitem{hubbard63}
J.~Hubbard.
\newblock Proc. R. Soc. London, Ser. A \textbf{276}, 238 (1963).

\bibitem{hubbard64a}
J.~Hubbard.
\newblock Proc. R. Soc. London, Ser. A \textbf{277}, 237 (1964).

\bibitem{hubbard64b}
J.~Hubbard.
\newblock Proc. R. Soc. London, Ser. A \textbf{281}, 401 (1964).

\bibitem{hubbard65}
J.~Hubbard.
\newblock Proc. R. Soc. London, Ser. A \textbf{285}, 542 (1965).

\bibitem{rasetti91}
M.~Rasetti, editor.
\newblock \emph{The Hubbard Model: Recent Results} (World Scientific,
  Singapore, 1991).

\bibitem{montorsi92}
A.~Montorsi, editor.
\newblock \emph{The Hubbard Model: A Collection of Reprints} (World Scientific,
  Singapore, 1992).

\bibitem{gebhard97}
F.~Gebhard, editor.
\newblock \emph{The Mott Metal-Insulator Transition: Model and Methods}
  (Springer, New York, 1997).

\bibitem{fazekas99}
P.~Fazekas.
\newblock \emph{Lecture Notes on Electron Correlation and Magnetism} (World
  Scientific, Singapore, 1999).

\bibitem{white89}
S.~R. White, D.~J. Scalapino, R.~L. Sugar, N.~E. Bickers, and R.~T. Scalettar.
\newblock Phys. Rev. B \textbf{39}, 839 (1989).

\bibitem{scalapino94}
D.~J. Scalapino.
\newblock In \emph{Proceedings of the International School of Physics, ``Enrico
  Fermi''}, edited by R.~A. Broglia and J.~R. Schrieffer (1994).
\newblock And references cited therein.

\bibitem{scalapino94b}
D.~J. Scalapino.
\newblock J. Low Temp. Phys. \textbf{95}, 169 (1994).

\bibitem{maier05}
T.~A. Maier, M.~Jarrell, T.~C. Schulthess, P.~R.~C. Kent, and J.~B. White.
\newblock Phys. Rev. Lett. \textbf{95}, 237001 (2005).

\bibitem{scalapino07}
D.~J. Scalapino.
\newblock \emph{Handbook of High Temperature Supercondutivity} (Springer, New
  York, 2007), chap.~13, pp. 495--526.

\bibitem{millis98}
A.~J. Millis.
\newblock Nature (London) \textbf{392}, 438 (1998).

\bibitem{white00}
S.~R. White and D.~J. Scalapino.
\newblock Phys. Rev. B \textbf{61}, 6320 (2000).

\bibitem{moreo05}
G.~Alvarez, M.~Mayr, A.~Moreo, and E.~Dagotto.
\newblock Phys. Rev. B \textbf{71}, 014514 (2005).

\bibitem{chang08}
C.-C. Chang and S.~Zhang.
\newblock Phys. Rev. B \textbf{78}, 165101 (2008).

\bibitem{reger88}
J.~D. Reger and A.~P. Young.
\newblock Phys. Rev. B \textbf{37}, 5978 (1988).

\bibitem{moriya85}
T.~Moriya.
\newblock \emph{Spin Fluctuations in Itinerant Electron Magnetism}, vol.~56 of
  \emph{Springer Series in Solid State Sciences} (Springer-Verlag, Berlin,
  1985).

\bibitem{moriya00}
T.~Moriya and K.~Ueda.
\newblock Adv. Phys. \textbf{49}, 555 (2000).

\bibitem{bickers89}
N.~E. Bickers and D.~J. Scalapino.
\newblock Ann. Phys. \textbf{193}, 206  (1989).

\bibitem{vilk97}
Y.~Vilk and A.-M. Tremblay.
\newblock J. Phys. I \textbf{7}, 1309 (1997).

\bibitem{borejsza04}
K.~Borejsza and N.~Dupuis.
\newblock Phys. Rev. B \textbf{69}, 085119 (2004).

\bibitem{schultz95}
H.~J. Schultz.
\newblock In \emph{The Hubbard Model: Its Physics and Mathematical Physics},
  edited by D.~Baeriswyl, D.~K. Campbell, J.~M. Carmelo, F.~Guinea, and
  E.~Louis (Springer, 1995).

\bibitem{demarco99}
B.~DeMarco and D.~S. Jin.
\newblock Science \textbf{285}, 1703 (1999).

\bibitem{truscott01}
A.~G. Truscott, K.~E. Strecker, W.~I. McAlexander, G.~B. Partridge, and R.~G.
  Hulet.
\newblock Science \textbf{291}, 2570 (2001).

\bibitem{ohara02}
K.~M. O'Hara, S.~L. Hemmer, M.~E. Gehm, S.~R. Granade, and J.~E. Thomas.
\newblock Science \textbf{298}, 2179 (2002).

\bibitem{kohl05}
M.~K\"ohl, H.~Moritz, T.~St\"oferle, K.~G\"unter, and T.~Esslinger.
\newblock Phys. Rev. Lett. \textbf{94}, 080403 (2005).

\bibitem{kohl06}
M.~K\"ohl and T.~Esslinger.
\newblock Europhys. News \textbf{37}, 18 (2006).

\bibitem{jordens08}
R.~Jordens, N.~Strohmaier, K.~Gunter, H.~Moritz, and T.~Esslinger.
\newblock Nature \textbf{455}, 204 (2008).

\bibitem{werner05}
F.~Werner, O.~Parcollet, A.~Georges, and S.~R. Hassan.
\newblock Phys. Rev. Lett. \textbf{95}, 056401 (2005).

\bibitem{vollhardt93}
D.~Vollhardt.
\newblock In \emph{Correlated Electron Systems}, edited by V.~J. Emery (World
  Scientific, Singapore, 1993), p.~57.

\bibitem{bulut94}
N.~Bulut, D.~J. Scalapino, and S.~R. White.
\newblock Phys. Rev. Lett. \textbf{73}, 748 (1994).

\bibitem{haas95}
S.~Haas, A.~Moreo, and E.~Dagotto.
\newblock Phys. Rev. Lett. \textbf{74}, 4281 (1995).

\bibitem{georges96}
A.~Georges, G.~Kotliar, W.~Krauth, and M.~J. Rozenberg.
\newblock Rev. Mod. Phys. \textbf{68}, 13 (1996).

\bibitem{preuss97}
R.~Preuss, W.~Hanke, C.~Gr\"ober, and H.~G. Evertz.
\newblock Phys. Rev. Lett. \textbf{79}, 1122 (1997).

\bibitem{grober00}
C.~Gr\"ober, R.~Eder, and W.~Hanke.
\newblock Phys. Rev. B \textbf{62}, 4336 (2000).

\bibitem{maier05rmp}
T.~Maier, M.~Jarrell, T.~Pruschke, and M.~H. Hettler.
\newblock Rev. Mod. Phys. \textbf{77}, 1027 (2005).

\bibitem{dare07}
A.-M. Dar\'{e}, L.~Raymond, G.~Albinet, and A.-M.~S. Tremblay.
\newblock Phys. Rev. B \textbf{76}, 064402 (2007).

\bibitem{blankenbecler81}
R.~Blankenbecler, D.~J. Scalapino, and R.~L. Sugar.
\newblock Phys. Rev. D \textbf{24}, 2278 (1981).

\bibitem{scalettar91}
R.~T. Scalettar, R.~M. Noack, and R.~R.~P. Singh.
\newblock Phys. Rev. B \textbf{44}, 10502 (1991).

\bibitem{jarrell09}
K. Michelsons and M. Jarrell, (unpublished).

\bibitem{quest09}
The development of this new determinant Quantum Monte Carlo program is part of
  the DOE SciDAC program. QUantum Electron Simulation Toolbox (QUEST) is a
  FORTRAN 90/95 package using new algorithms, such as delayed updating, and
  integrating modern BLAS/LAPACK numerical kernels. QUEST has integrated
  several legacy codes by modularizing their computational components for ease
  of maintenance and program interfacing. QUEST also allows general lattice
  geometries. The current version can be accessed via
  \url{http://www.cs.ucdavis.edu/~bai/QUEST}.

\bibitem{hubsims2}
A useful collection of lectures summarizing some of the alternate approaches to
  Hubbard model QMC is contained in the NATO Advanced Study Institute: Quantum
  Monte Carlo Methods in Physics and Chemistry (1998), available at
  \url{http://www.phys.uri.edu/~nigh/QMC-NATO/webpage/abstracts/lecturers.html%
}. These include ``Phase Separation in the 2D Hubbard Model: A Challenging
  Application of Fixed-Node QMC'' (Bachelet); ``Quantum Monte Carlo for Lattice
  Fermions,'' (Muramatsu); and ``Constrained Path Quantum Monte Carlo for
  Fermions,'' (Zhang).

\bibitem{vanhoucke08}
K.~Van~Houcke, E.~Kozik, N.~V. Prokof\'ev, and B.~V. Svistunov.
\newblock In \emph{Computer Simulation Studies in Condensed Matter Physics
  XXI}, edited by D.~P. Landau, S.~P. Lewis, and H.~B. Sch\"uttler
  (Springer-Verlag, Berlin, 2008).

\bibitem{prokofev08}
N.~V. Prokof'ev and B.~V. Svistunov.
\newblock Phys. Rev. B \textbf{77}, 125101 (2008).

\bibitem{hirsch85}
J.~E. Hirsch.
\newblock Phys. Rev. B \textbf{31}, 4403 (1985).

\bibitem{loh90}
E.~Y. Loh, J.~E. Gubernatis, R.~T. Scalettar, S.~R. White, D.~J. Scalapino, and
  R.~L. Sugar.
\newblock Phys. Rev. B \textbf{41}, 9301 (1990).

\bibitem{huse88}
D.~A. Huse.
\newblock Phys. Rev. B \textbf{37}, 2380 (1988).

\bibitem{hirsch89}
J.~E. Hirsch and S.~Tang.
\newblock Phys. Rev. Lett. \textbf{62}, 591 (1989).

\bibitem{sandvik97}
A.~W. Sandvik.
\newblock Phys. Rev. B \textbf{56}, 11678 (1997).

\bibitem{delannoy05}
J.-Y.~P. Delannoy, M.~J.~P. Gingras, P.~C.~W. Holdsworth, and A.-M.~S.
  Tremblay.
\newblock Phys. Rev. B \textbf{72}, 115114 (2005).

\bibitem{schrieffer89}
J.~R. Schrieffer, X.~G. Wen, and S.~C. Zhang.
\newblock Phys. Rev. B \textbf{39}, 11663 (1989).

\bibitem{feynman82}
R.~P. Feynman \textbf{21}, 467 (1982).

\end{thebibliography}

\end{document}